\documentclass[twocolumn,showpacs,preprintnumbers,amsmath,amssymb,superscriptaddress,aps,pra,10pt]{revtex4-1}

\usepackage{color}
\usepackage{graphicx}
\usepackage{citesort}

\newcommand{\comment}[1]{}

\newcommand{\ie}{\mbox{i.e.\ }}
\newcommand{\eg}{\mbox{e.g.\ }}

\newcommand{\eqnref}[1]{Eq.~(\ref{#1})}

\newcommand{\figref}[1]{Fig.~\ref{#1}}
\newcommand{\secref}[1]{Section~\ref{#1}}

\newcommand{\eps}{\varepsilon}

\newcommand{\Power}{{\mathcal{P}}}

\newcommand{\total}{{\rm{d}}}
\newcommand{\imag}{{\rm{i}}}

\newcommand{\AmplGain}{\tilde \PowerGain}
\newcommand{\PowerGain}{\Gamma}
\newcommand{\TotalGain}{\mathcal{A}_{\text{dB}}}
\newcommand{\FOM}{\mathcal{F}}

\newcommand{\cc}{{\rm{c.c.}}}

\renewcommand{\vec}[1]{{\bf{#1}}}

\newcommand{\dirderiv}{\partial}

\newcommand{\mode}[2]{{#1}^{(#2)}}

\newcommand{\BSBS}{{\text{B}}}
\newcommand{\FSBS}{{\text{F}}}
\newcommand{\TPA}{{\text{2PA}}}
\newcommand{\FCA}{{\text{FCA}}}
\newcommand{\linear}{{\text{lin}}}
\newcommand{\loss}{{\text{loss}}}

\newcommand{\eff}{{\text{eff}}}
\newcommand{\opt}{{\text{opt}}}

\newcommand{\acoustic}{\text{ac}}
\newcommand{\optical}{\text{opt}}
\newcommand{\acaverage}[1]{\Big \langle \ {#1} \ \Big \rangle_{T_\acoustic}}
\newcommand{\optaverage}[1]{\Big \langle \ {#1} \ \Big \rangle_{T_\optical}}

\begin{document}

\title{Impact of nonlinear loss on Stimulated Brillouin Scattering}
\author{C. Wolff}
\affiliation{
  Centre for Ultrahigh bandwidth Devices for Optical Systems (CUDOS), 
}
\affiliation{
  School of Mathematical and Physical Sciences, University of Technology Sydney, NSW 2007, Australia
}

\author{P. Gutsche}
\affiliation{
  School of Mathematical and Physical Sciences, University of Technology Sydney, NSW 2007, Australia
}

\author{M.~J. Steel}
\affiliation{
  Centre for Ultrahigh bandwidth Devices for Optical Systems (CUDOS), 
}
\affiliation{
  MQ Photonics Research Centre, Department of Physics and Astronomy, Macquarie University Sydney, NSW 2109, Australia
}

\author{B.~J. Eggleton}
\affiliation{
  Centre for Ultrahigh bandwidth Devices for Optical Systems (CUDOS),
}
\affiliation{
  Institute of Photonics and Optical Science (IPOS), School of Physics,
  University of Sydney, NSW 2006, Australia
}

\author{C.~G. Poulton}
\affiliation{
  Centre for Ultrahigh bandwidth Devices for Optical Systems (CUDOS),
}
\affiliation{
  School of Mathematical and Physical Sciences, University of Technology Sydney, NSW 2007, Australia
}

\email{christian.wolff@uts.edu.au}

\date{\today}

\begin{abstract}
  We study the impact of two-photon absorption (2PA) and fifth-order nonlinear 
  loss such as 2PA-induced free-carrier absorption in semiconductors on the 
  performance of Stimulated Brillouin Scattering devices.
  We formulate the equations of motion including effective loss coefficients,
  whose explicit expressions are provided for numerical evaluation in any
  waveguide geometry.
  We find that 2PA results in a monotonic, algebraic relationship between 
  amplification, waveguide length and pump power, whereas fifth-order losses
  lead to a non-monotonic relationship.
  We define a figure of merit for materials and waveguide designs in the
  presence of fifth-order losses.
  From this, we determine the optimal waveguide length for the case of 2PA
  alone and upper bounds for the total Stokes amplification for the case of
  2PA as well as fifth-order losses.
  The analysis is performed analytically using a small-signal approximation
  and is compared to numerical solutions of the full nonlinear modal equations.
\end{abstract}

\maketitle



\section{Introduction}

In recent years the field of opto-mechanics has broadened from quantum-opto-mechanical 
research undertaken in high-Q resonators~\cite{Kippenberg2008} to include the interaction of light 
with vibrations in high index-contrast optical waveguides ~\cite{Eggleton2013}. 
The dominant opto-mechanical effect to occur in waveguiding geometries is
Stimulated Brillouin Scattering (SBS), which is the scattering of light from the travelling 
grating that is formed by an acoustic wave in the optical medium~\cite{Boyd2003}.
SBS was first proposed by Brillouin~\cite{Brillouin1922} and subsequently observed
in various systems, ranging from first experiments with quartz~\cite{Chiao1964}, 
over the field of fibre optics, where it is well-known as a strong third-order
nonlinearity~\cite{Agrawal} to more recent studies in on-ship waveguide such as
chalcogenide rib waveguides~\cite{Pant2011}.
This evolution has led naturally to the most recent investigation of SBS in 
silicon nanowires~\cite{Shin2013,vanLaer2015} based on exciting theoretical 
work~\cite{Rakich2012} in the past few years. 
The very strong acousto-optical interaction that can be achieved in this system
provides the potential to implement a number of established SBS-applications
in on-chip platforms; this includes novel light 
sources~\cite{Kabakova2013,Hu2014,Buettner2014}, non-reciprocal light 
propagation~\cite{Huang2011,Aryanfar2014}, slow light~\cite{Thevenaz2008},
and signal processing in the context of microwave 
photonics~\cite{Vidal2007,Li2013,Morrison2014}.

Conventionally, the description of SBS is based on the approximation that SBS
is by far the dominating nonlinear effect.
A consequence of this approximation is that the total amplification of the 
Stokes wave along the waveguide's full length should be proportional to the 
power of the injected pump beam.
The Stokes wave should then initially exhibit exponential growth until it 
starts to deplete the pump; additional Stokes power can be gained by using a 
longer waveguide or by increasing the pump power. 
However, nonlinear loss is known to have relevant impact on the related effect
of stimulated Raman-scattering~\cite{Rukhlenko2010} (especially in silicon
photonics) and recent experiments in silicon waveguides at telecom wavelengths
have also strongly suggested that nonlinear loss has an appreciable impact on 
the overall dynamic of the SBS process~\cite{vanLaer2015}.
In this context, third-order loss -- \ie two-photon absorption (2PA) -- can 
be expected to have a qualitatively different effect on the SBS gain compared 
to fifth-order processes such as three-photon absorption (3PA) and 2PA-induced 
free carrier absorption (FCA).
The latter is of considerable importance for semiconductor waveguides, which 
are the best candidates for affordable highly integrated optical circuits. 
Evaluating the impact of nonlinear loss terms on the SBS gain is critical 
for the design of future SBS-based devices. 

In this paper we study the impact of 2PA and 2PA-induced FCA on the performance
of SBS. 
To this end, we derive nonlinear loss coefficients and solve the SBS-equations
analytically within a small-signal approximation.
We thereby deviate from the Raman literature~\cite{Rukhlenko2010}, which is 
necessary because SBS (in contrast to Raman scattering) is not always a 
forward-process, which would lead to directly integrable differential equations.
Our analytical solutions provide strict upper bounds for key quantities such as
output powers and amplification, which we express in terms of figures of merit 
for SBS in the presence of different types of nonlinear loss.
For the case of negligible fifth-order losses (i.e. where FCA can be neglected), 
we find 
\begin{align}
  \FOM_\TPA = \frac{\PowerGain}{2\beta},
\end{align}
whereas the appropriate figure of merit for the case of dominant 2PA-induced FCA 
with a weak 2PA-perturbation is
\begin{align}
  \FOM_\FCA = \frac{\PowerGain - 2\beta}{2\sqrt{\alpha \gamma}},
  \label{eqn:intro_fom_fca}
\end{align}
where $\alpha$, $\beta$, $\gamma$ and $\PowerGain$ are the effective linear 
loss coefficient, effective 2PA-parameter, effective fifth-order loss 
(\eg 2PA-induced FCA) parameter and the SBS-gain, respectively.
These figures of merit must be greater than $1$ in order for the Stokes wave to 
be amplified. 
The application of these figures of merit can result in upper bounds for the 
Stokes amplification; an investigation of the specific limits resulting from the 
presence of free carriers has been submitted 
separately as a rapid communication~\cite{Wolff2015b}.

The layout of the manuscript is as follows: in \secref{sec:preliminaries}, 
we state the preliminaries of our analysis, we
introduce the relevant equations of motion, define and apply the small-signal
approximation for the Stokes wave and state the expressions for the effective 
nonlinear loss coefficients for the case of intra-mode forward and backward SBS.
In \secref{sec:solutions}, we solve the small-signal equations analytically for 
systems that exhibit 2PA and linear loss, for systems with 2PA-induced FCA
with and without linear loss and we discuss the leading order perturbative
expression for case of linear loss and FCA with additional weak 2PA.
Based on these analytical solutions we proceed in \secref{sec:design} to derive 
figures of merit and present design guidelines for maximising SBS gain 
in arbitrary waveguide geometries. Conclusions and implications are discussed  
in \secref{sec:conclusion}. Finally, we include three
Appendices, in which we state the nonlinear coefficients for the more general
case of inter-mode SBS as well as their derivations.

\section{Preliminaries and approximations}
\label{sec:preliminaries}

We consider the interaction between optical and acoustic fields in waveguides 
with a material cross 
section that is invariant along the $z$-axis and supports both optical and
acoustic guided modes.
For the materials we assume the absence of magnetic response ($\mu_r = 1$), of
material dispersion, of even-order nonlinearities (in particular 
piezoelectricity) and of the Kerr effect.
We explicitly include weak linear and odd-order nonlinear optical loss.
The formulation is based on our previous work on the theory of SBS in integrated
waveguides~\cite{Wolff2015a}.
We now introduce the equations and approximations
that are solved throughout the remainder of this paper.
We use SI-units throughout.

\subsection{Local acoustic approximation, power equations}

The starting point of this paper are the stationary coupled mode equations for 
the optical Stokes amplitude $\mode{a}{1}(z)$ at angular frequency $\mode{\omega}{1}$, 
the optical pump amplitude 
$\mode{a}{2}(z)$ at angular frequency $\mode{\omega}{2}=\mode{\omega}{1}+\Omega$,
and the acoustic amplitude $b(z)$:
\begin{widetext}
\begin{align}
  \dirderiv_z \mode{a}{1} 
  \, + \, \left(\tilde \alpha_1 + \tilde \beta_{11} |\mode{a}{1}|^2 
  + \tilde \gamma_{111} |\mode{a}{1}|^4\right) \, \mode{a}{1} 
  = & 
  - \left(2 \tilde \beta_{12} + 4 \tilde \gamma_{112} |\mode{a}{1}|^2 
  + \tilde \gamma_{122} |\mode{a}{2}|^2\right) \, |\mode{a}{2}|^2 \mode{a}{1}
  \, - \, \frac{\imag \mode{\omega}{1} Q}{\mode{\Power}{1}} \mode{a}{2} b^\ast,
  \label{eqn:sbs_stokes}
  \\
  \dirderiv_z \mode{a}{2} 
  \, + \, \left(\tilde \alpha_2 + \tilde \beta_{22} |\mode{a}{2}|^2 
  + \tilde \gamma_{222} |\mode{a}{2}|^4 \right) \, \mode{a}{2} 
  = & 
  - \left(2 \tilde \beta_{21} + 4 \tilde \gamma_{221} |\mode{a}{2}|^2 
  + \tilde \gamma_{211} |\mode{a}{1}|^2 \right) \, |\mode{a}{1}|^2 \mode{a}{2}
  \, - \, \frac{\imag \mode{\omega}{2} Q^\ast}{\mode{\Power}{2}} \mode{a}{1} b,
  \label{eqn:sbs_pump}
  \\
  \dirderiv_z b + \alpha_b b 
  = &
  -\frac{\imag \Omega Q_b}{\Power_b} [\mode{a}{1}]^\ast \mode{a}{2},
  \label{eqn:sbs_ac}
\end{align}
\end{widetext}
where the acousto-optic coupling parameter $Q$ and the respective (signed) 
modal power fluxes $\mode{\Power}{1,2}$ appear as introduced in Ref.~\cite{Wolff2015a}.
Note that the modal powers can be negative for modes that travel backwards 
(\ie in the negative $z$-direction) and that the acousto-optic coupling is 
real-valued in the absence of loss.
The quantities
$\tilde \alpha_i$, $\tilde \beta_{ij}$ and $\tilde \gamma_{ijk}$ are the 
modal linear and nonlinear loss coefficients as derived in the Appendices of 
this manuscript.
They are explicitly stated in Eqs.~(\ref{eqn:appx_linear}--\ref{eqn:appx_fca}).
Note the nontrivial factors $2$ and $4$ in front of the terms involving
$\tilde \beta_{12}$, $\tilde \beta_{21}$, $\tilde \gamma_{112}$ and
$\tilde \gamma_{221}$.

As a first simplification, we assume that the acoustic decay length $\alpha_b$ 
is much smaller than the length scale on which the optical envelopes vary.
This is a valid assumption for SBS in fibres and in moderately long (mm-scale) integrated 
waveguides~\cite{Pant2011}.
As a result, we can approximate the acoustic profile:
\begin{align}
  b(z) \approx - \frac{\imag \Omega Q [\mode{a}{1}(z)]^\ast \mode{a}{2}(z)}{\alpha_b \Power_b}.
  \label{eqn:sbs_ac_approx}
\end{align}
Under the assumption of phase matching and in conjunction with the approximation 
$\mode{\omega}{1} \approx \mode{\omega}{2} = \omega$, this allows us to eliminate 
the acoustic envelope in the equations of motion by inserting the substitutions 
\begin{align}
  - \frac{\imag \omega Q \mode{a}{2} b^\ast}{\mode{\Power}{1}}
  \approx &
  \AmplGain^\ast \mode{\Power}{2} |\mode{a}{2}|^2 \mode{a}{1},
  \\
  - \frac{\imag \omega Q \mode{a}{1} b}{\mode{\Power}{2}}
  \approx &
  -\AmplGain \mode{\Power}{1} |\mode{a}{1}|^2 \mode{a}{2},
  \intertext{with}
  \AmplGain = & \frac{\omega \Omega |Q|^2}{\mode{\Power}{1} 
  \mode{\Power}{2} \Power_b \alpha_b}
\end{align}
into Eqs.~(\ref{eqn:sbs_stokes}, \ref{eqn:sbs_pump}).
The coupling parameter ceases to be real-valued if phase-matching is 
broken~\cite{Wolff2015a}.

Next, we introduce the expression for the power fluxes in the Stokes and the pump
modes and their derivatives along the waveguide:
\begin{align}
  \mode{P}{1,2} = & \mode{\Power}{1,2} |\mode{a}{1,2}|^2,
  \\
  \partial_z \mode{P}{1,2} = & 
  2 \mode{\Power}{1,2} [\mode{a}{1,2}]^\ast \ \partial_z \mode{a}{1,2}.
\end{align}
By inserting these expressions into 
Eqs.~(\ref{eqn:sbs_stokes},\ref{eqn:sbs_pump}), we obtain the equations of 
motion for the optical powers in a local acoustic approximation:
\begin{widetext}
\begin{align}
  \dirderiv_z \mode{P}{1} 
  \, + \, \left(\alpha_1 + \beta_{11} \mode{P}{1}
  + \gamma_{111} [\mode{P}{1}]^2 \right) \, \mode{P}{1} 
  = & 
  - \left(2 \beta_{12} - \PowerGain + 4 \gamma_{112} \mode{P}{1}
  + \gamma_{122} \mode{P}{2}\right) \, \mode{P}{1} \mode{P}{2}
  \label{eqn:sbs_stokes_pwr}
  \\
  \dirderiv_z \mode{P}{2} 
  \, + \, \left(\alpha_2 + \beta_{22} \mode{P}{2}
  + \gamma_{222} [\mode{P}{2}]^2 \right) \, \mode{P}{2} 
  = & 
  - \left(2 \beta_{21} + \PowerGain + 4 \gamma_{221} \mode{P}{2}
  + \gamma_{211} \mode{P}{1} \right) \, \mode{P}{1} \mode{P}{2},
  \label{eqn:sbs_pump_pwr}
\end{align}
\begin{center}
  with the power-related coefficients
\end{center}
\begin{align}
  \alpha_i = & 2 \tilde \alpha_i,
  &
  \beta_{ij} = & \frac{2 \tilde \beta_{ij}}{\mode{\Power}{j}},
  &
  \gamma_{ijk} = & \frac{2 \tilde \gamma_{ijk}}{\mode{\Power}{j}\mode{\Power}{k}},
  &
  \PowerGain = & 2 \Re\{\AmplGain\}.
\end{align}
\end{widetext}

\subsection{Small-signal approximation}

We now introduce the central approximation of our work: 
We assume that at every position $z$ inside the waveguide, the Stokes wave is
much weaker than the pump wave:
\begin{align}
  \big|\mode{P}{1}(z)\big| \ll \big|\mode{P}{2}(z)\big|.
\end{align}
As a consequence, a number of terms can be dropped from the 
Eqs.~(\ref{eqn:sbs_stokes_pwr},\ref{eqn:sbs_pump_pwr}):
all terms that are of at least second order in $\mode{P}{1}$
in \eqnref{eqn:sbs_stokes_pwr} and all terms that involve $\mode{P}{1}$ in 
\eqnref{eqn:sbs_pump_pwr}.
As a result, we obtain a simpler set of differential equations; one separable 
equation and one which depends on only a single quantity:
\begin{align}
\label{eq:governing}
  \nonumber
  \dirderiv_z \mode{P}{1} = & -\alpha_1 \mode{P}{1}
  \\
  &
  + (\PowerGain - 2 \beta_{12} - \gamma_{122} \mode{P}{2}) \mode{P}{2} \mode{P}{1},
  \\
  \dirderiv_z \mode{P}{2} = &  
  -(\alpha_2 + \beta_{22} \mode{P}{2} + \gamma_{222} [\mode{P}{2}]^2) \mode{P}{2}.
\end{align}
At first, this may seem to be a severe approximation.
However, it can be motivated as follows: First, the approximation leads to a distribution 
of the Stokes power along the waveguide that
is strictly proportional to the injected Stokes power.
This means that the SBS-active waveguide acts as a linear amplifier for the
Stokes signal and only in this situation can the waveguide be expressed by an 
amplification factor that does not explicitly depend on the Stokes power.
More importantly, this provides an upper bound for the amplification that
is realisable for a certain set of parameters $\alpha, \beta, \gamma$ and 
$\PowerGain$, because every term we neglect in the small-signal approximation
introduces further loss.
In other words, the Stokes amplification predicted within this approximation is 
a upper bound for the amplification that can be observed in reality with 
a finite Stokes input power.
Stokes amplification can be observable in an experiment if and only if the 
small-signal approximation predicts it.

\subsection{Special cases: intra-mode SBS }

The expressions derived thus far apply to both inter-mode and intra-mode SBS,
both in forward and backward configuration determined by the sign of the modal 
normalisation power $\mode{\Power}{1}$  of the Stokes wave (positive for forward 
scattering, negative for backward scattering).
We now restrict ourselves to the more common case of interaction within the 
same branch of the optical dispersion relation, \ie we assume either intra-mode 
forward SBS (FSBS) or backward SBS (BSBS).
The analysis presented in the remainder of this paper can be carried out without
this simplification and yields results of the same form, yet less transparent 
due to the required mode labels.
In the case of FSBS, the Stokes and pump mode and respective normalisation power 
are identical:
\begin{align}
  \mode{\widetilde{\vec e}}{1}(x, y) 
  = & \mode{\widetilde{\vec e}}{2}(x, y)
  = \widetilde{\vec e}(x, y) 
  \\
  \mode{\Power}{1} = & \mode{\Power}{2} =  \Power.
\end{align}
If this is inserted in the expressions for the loss coefficients derived in the
Appendices, we find that all coefficients of equal order are identical:
\begin{align}
  \nonumber
  \alpha = & \alpha_1 = \alpha_2 
  \\
  = &
  \frac{2 \eps_0 \omega}{\Power} \int \total^2 r \ 
    | \widetilde{\vec e} |^2 \Im\{\eps_r\},
  \\
  \nonumber
  \beta = & \beta_{11} = \beta_{12} = \beta_{21} = 
  \beta_{22}
  \\
  = & \frac{2}{\Power^2} \int \total^2 r \ \big(
  2 | \widetilde{\vec e} \cdot \widetilde{\vec e}^\ast |^2 + 
  | \widetilde{\vec e} \cdot \widetilde{\vec e} |^2 \big)
  \Sigma^\TPA,
  \\
  \nonumber
  \gamma = & \gamma_{111} = \gamma_{112} = 
  \gamma_{122} = \gamma_{211} = \gamma_{221} = 
  \gamma_{222}
  \\
  = & \frac{2}{\Power^3}
    \int \total^2 r \ |\widetilde{\vec e}|^2
    \Big[|\widetilde{\vec e} \cdot \widetilde{\vec e}|^2 
    + 2 (|\widetilde{\vec e}|^2)^2 \Big]
    \Sigma^\FCA,
\end{align}
where $\Sigma^\TPA$ and $\Sigma^\FCA$ are nonlinear conductivities of the 
material associated with 2PA and 2PA-induced FCA (see the Appendices for 
explicit expressions for values for the case of silicon).
In the case of BSBS, the Stokes and pump mode are identical up to complex
conjugation and the modal normalisation powers differ only in their sign:
\begin{align}
  \mode{\widetilde{\vec e}}{1}(x, y) = & [\mode{\widetilde{\vec e}}{2}]^\ast(x, y) 
  = \widetilde{\vec e}(x, y),
  \\
  \mode{\Power}{1} = & -\mode{\Power}{2} = -\Power.
\end{align}
As it turns out, the resulting loss coefficients like the modal powers are
identical up to a sign. 
We find for the loss coefficients:
\begin{align}
  \alpha_1 = & -\alpha, 
  & \alpha_2 = & \alpha,
  \\
  \beta_{1i} = & -\beta, 
  & \beta_{2i} = & \beta,
  \\
  \gamma_{1ij} = & - \gamma,
  & \gamma_{2ij} = & \gamma,
\end{align}
for all $i$ and $j$, where the expressions for $\alpha$, $\beta$ and $\gamma$ 
are the same as given for the case of FSBS.

\section{Analytical solutions}
\label{sec:solutions}

In this section, we present and discuss the analytical solutions to the 
small-signal equations for several experimentally relevant combinations of
loss mechanisms.
As a first step, we restate the equations of motion for the case of backward
SBS. 
For convenience, we prefer the Stokes power to be positive definite, \ie
$\mode{P}{1,\BSBS} = \Power |\mode{a}{1}|^2$, leading to the equations:
\begin{align}
  \nonumber
  \dirderiv_z \mode{P}{1,\BSBS} = & \alpha \mode{P}{1,\BSBS}
  - 
  \\
  & \ 
  (\PowerGain - 2 \beta - \gamma \mode{P}{2,\BSBS}) \mode{P}{2} \mode{P}{1,\BSBS},
  \label{eqn:stokes_start_bsbs}
  \\
  \dirderiv_z \mode{P}{2} = &  
  -(\alpha + \beta \mode{P}{2} + \gamma [\mode{P}{2}]^2) \mode{P}{2}.
  \label{eqn:pump_start}
\end{align}
The waveguide is assumed to extend from $z=0$ to $z=L$.
SBS-gain is usually specified in units of decibel per unit length of waveguide.
However, this assumes that the Stokes wave grows exponentially, which is not  
the case in the presence of nonlinear loss.
Thus, the natural quantity to compare is the total Stokes amplification in 
decibel over a waveguide of finite length $L$:
\begin{align}
  \TotalGain^\BSBS(L) = 10 \log_{10} \left[
  \frac{\mode{P}{1,\BSBS}(0)}{\mode{P}{1,\BSBS}(L)} \right]
\end{align}

For the case of forward SBS with $\mode{P}{1,\FSBS} \geq 0$, only the sign 
of the right hand side of \eqnref{eqn:stokes_start_bsbs} flips:
\begin{align}
  \nonumber
  \dirderiv_z \mode{P}{1,\FSBS} = & - \alpha \mode{P}{1,\FSBS}
  \\
  & + (\PowerGain - 2 \beta - \gamma \mode{P}{2}) \mode{P}{2} \mode{P}{1,\FSBS},
  \label{eqn:fsbs_stokes_start}
\end{align}
As a result, the forward SBS solution is exactly the inverse of the
corresponding backward SBS solution apart from a constant factor $C$:
\begin{align}
  \mode{P}{1,\FSBS}(z) = \frac{C}{\mode{P}{1,\BSBS}(z)}.
\end{align}
The total forward amplification is thus identical to the backward Stokes 
amplification:
\begin{align}
  \TotalGain^\FSBS(L) = & 10 \log_{10} \left[
  \frac{\mode{P}{1,\FSBS}(L)}{\mode{P}{1,\FSBS}(0)} \right]
  = \TotalGain^\BSBS(L).
\end{align}
For this reason, we can focus on either of the two types of SBS whenever we 
employ the small-signal approximation.
This is the case for the remainder of this paper except for the discussion
in \secref{sec:ss_fca_and_lin}, where we compare our analytical expressions 
to numerical solutions of the large-signal 
Eqs.~(\ref{eqn:sbs_stokes},\ref{eqn:sbs_pump},\ref{eqn:sbs_ac_approx}).
We choose to base the following discussions on the equations for FSBS and will 
omit the label $\FSBS$ where possible:
\begin{align}
  \dirderiv_z \mode{P}{1} = & - \alpha \mode{P}{1}
  + (\PowerGain - 2 \beta - \gamma \mode{P}{2}) \mode{P}{2} \mode{P}{1},
  \label{eqn:stokes_start}
  \\
  \dirderiv_z \mode{P}{2} = &  
  -(\alpha + \beta \mode{P}{2} + \gamma [\mode{P}{2}]^2) \mode{P}{2}.
  \label{eqn:stokes_pump}
\end{align}
The general approach is as follows: we first solve the equation for the
pump field, then insert that solution into the equation for Stokes field and
integrate the resulting differential equation.

\subsection{SBS with 2PA and linear loss}
\label{sec:ss_tpa_and_lin}

\begin{figure}[t]
  \includegraphics[width=\columnwidth]{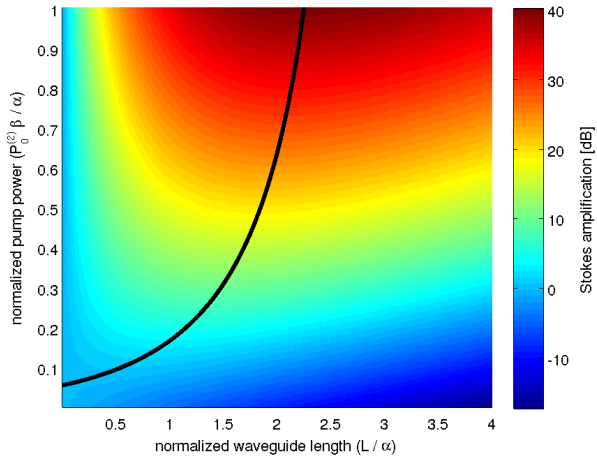}
  \caption{
    Stokes amplification in decibel of a waveguide that is subject to linear
    loss and 2PA as a function of the pump power level and the waveguide 
    length in natural units for a SBS gain of $\PowerGain = 20 \beta$.
    Note that for every fixed pump power above a certain threshold, the 
    amplification grows with increasing waveguide length, reaches a maximum and 
    then drops off again.
    These maxima indicate optimal waveguide lengths and are connected with a
    solid black line.
  }
  \label{fig:ss_tpa_gain}
\end{figure}

First, we investigate the case that the material does not exhibit fifth-order
loss.
This applies for example to glasses, including the chalcogenide rib waveguides 
that are the basis of recent on-chip SBS activities~\cite{Eggleton2013}, or
to semiconductors illuminated below the half-bandgap.
This case implies $\{\alpha, \beta\} \neq 0$; $\gamma = 0$.
The resulting equation for the pump power is:
\begin{align}
  \dirderiv_z \mode{P}{2} + \alpha \mode{P}{2} + \beta [\mode{P}{2}]^2 = 0.
  \label{eqn:bsbs_tpa_lin_pumpstart}
\end{align}
This is an equation of the Riccati type and can therefore be solved using the
ansatz $ \mode{P}{2}(z) = y' / (\beta y)$, resulting in the solution
\begin{align}
  \mode{P}{2}(z) = & \frac{\alpha \exp(-\alpha z)}{\beta [\zeta - \exp(-\alpha z)]},
  \\
  \text{where} \quad \zeta = & 1 + \frac{\alpha}{\beta \mode{P}{2}_0}
  \label{eqn:bsbs_tpa_and_lin_pump}
\end{align}
parametrises the input pump power.
Next, we insert this result in the Stokes equation:
\begin{align}
  \dirderiv_z \mode{P}{1} = & -\alpha 
  \left[ 1 - \frac{(\PowerGain - 2 \beta) \exp(-\alpha z)}{\beta[\zeta - \exp(-\alpha z)]} \right]
  \mode{P}{1},
\end{align}
which can be readily solved:
\begin{align}
  \frac{\dirderiv_z \mode{P}{1}}{\mode{P}{1}} = &
  -\alpha \frac{\PowerGain - 2 \beta}{\beta} \cdot
  \frac{\alpha \exp(-\alpha z)}{\zeta - \exp(-\alpha z)};
  \\
  \mode{P}{1}(z) = & S^{\TPA} \exp(-\alpha z) 
  \left[ \zeta - \exp (-\alpha z) \right]^{(\PowerGain / \beta - 2)},
  \label{eqn:ss_stokes_tpa_and_lin}
  \intertext{where}
  S^{\TPA} = & \mode{P}{1}_0 ( \zeta - 1 )^{(2 - \PowerGain / \beta)}
\end{align}
parametrises the input Stokes power $\mode{P}{1}_0 = \mode{P}{1}(0)$.
From \eqnref{eqn:ss_stokes_tpa_and_lin}, we can readily derive the Stokes 
amplification in the presence of 2PA and linear loss:
\begin{align}
  \TotalGain^{\TPA+\linear}(L)
  = & \frac{10}{\ln 10} \Bigg\{ - \alpha L 
    + \left(\PowerGain - 2 \beta \right) \times
  \label{eqn:ss_gain_tpa_and_lin}
    \\
  \nonumber
    & \quad 
  \left[ \ln \left(
  1 + \beta \mode{P}{2}_0 [1 - \exp(-\alpha L)] / \alpha
  \right)
  \right]^{1/\beta}
\Bigg\}.
\end{align}
This result comprises two parts: First the exponential decay of the Stokes
power due to the linear loss $\alpha L$.
Second, the effect of SBS and 2PA.
The logarithm is always non-zero and can be interpreted so that 2PA and 
linear loss reduce the SBS-effective waveguide length and SBS-effective pump
power; it grows monotonically as a function of the waveguide length $L$ and the pump
power $\mode{P}{2}_0$, but saturates for $\alpha L \gg 1$ while being 
unbounded with respect to $\mode{P}{2}_0$.
The prefactor $(\PowerGain - 2 \beta)$ finally is the central result of this 
analysis.
It states that the SBS-gain of a waveguide with 2PA present is effectively
reduced by twice the 2PA-coefficient: $\PowerGain_\eff = \PowerGain - 2 \beta$.
Thus, SBS can only be observed if $\PowerGain / (2 \beta) > 1$.
Therefore, although any level of total Stokes amplification can be obtained by
increasing the pump power, increasing the waveguide length much beyond
$1/\alpha$ is not useful.
To illustrate this, the general dependence of $\TotalGain^{(\TPA+\linear)}$ 
for an arbitrary chosen ratio between SBS gain and 2PA-coefficient of
$\PowerGain = 20 \beta$ is depicted in \figref{fig:ss_tpa_gain} over a wide
range of waveguide lengths and pump powers.

To conclude, we consider the case of vanishing linear loss 
$\alpha \rightarrow 0$.
To this end, we approximate $\exp(-\alpha L) \approx 1 - \alpha L$ in
\eqnref{eqn:ss_stokes_tpa_and_lin} and \eqnref{eqn:ss_gain_tpa_and_lin}
to find:
\begin{align}
  \TotalGain^{\TPA}(L)
  = & \frac{10}{\ln 10} \left(\frac{\PowerGain}{\beta} - 2\right) \ln \left(
  1 + \beta L \mode{P}{2}_0 \right)
  \label{eqn:ss_gain_tpa}
  \\
\intertext{with the corresponding Stokes amplitude}
\mode{P}{1}(z) = & \mode{P}{1}_0 
  \left[ 1 + z \beta \mode{P}{2}_0 \right]^{(\PowerGain / \beta - 2)}.
\end{align}
We state this result to point out that in the case of 2PA as the only loss
mechanism, the solution is algebraic rather than transcendental and that any 
arbitrary exponent can be realised.
For example, it is possible to connect the Stokes power linearly or 
quadratically to the pump power by choosing $\PowerGain = 3 \beta$ or 
$\PowerGain = 4 \beta$, respectively.
We cannot think of any useful application for this analog computation, but 
find it an amusing curiosity.

\subsection{SBS with FCA alone}
\label{sec:ss_fca}

\begin{figure}[t]
  \includegraphics[width=0.96\columnwidth]{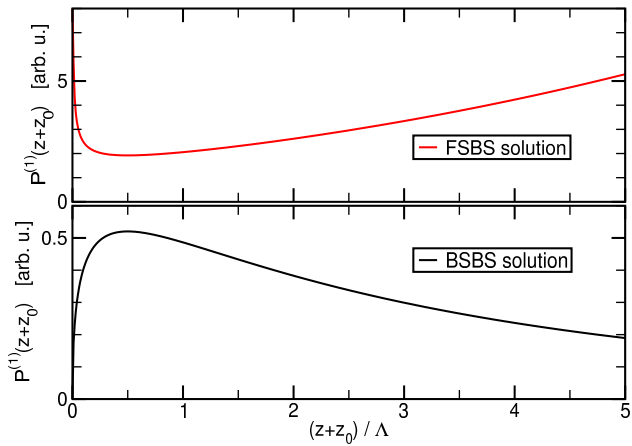}
  \caption{
    Solution to the problem of SBS with free carrier absorption in the
    forward SBS (top panel) and backward SBS (bottom panel) configuration
    for the solution normalisation constant set to $S^\FCA = \sqrt{\Lambda}$.
    Note the respective extrema of the solutions at $z + z_0 = \Lambda/2$.
    \label{fig:sbs_fca_stokes}
  }
\end{figure}

Next, we focus on the case $\{\alpha, \beta\} = 0$; $\gamma \neq 0$, which 
describes a system subject only to 2PA-induced free carrier
absorption, but neither linear loss nor 2PA itself:
2PA is assumed to generate carriers but not absorb a noticeable amount of 
energy itself.
The following discussion is intended as a preparation for the discussion of
2PA-induced FCA in combination with linear loss in \secref{sec:ss_fca_and_lin}
and for the qualitative discussion in \secref{sec:ss_fca_lin_and_weak_tpa}.
The results can also be directly applied to situations where both $\alpha$ and
$\beta$ are very small, \eg for very good semiconductor waveguides.

Within this section, the pump power satisfies the equation
\begin{align}
  \dirderiv_z \mode{P}{2} = -\gamma [\mode{P}{2}]^3,
\end{align}
which has the solution
\begin{align}
  \mode{P}{2}(z) = & \frac{1}{\sqrt{ 2 \gamma (z + z_0)}},
  \label{eqn:fca_pump}
  \\
  \text{where} \quad
  z_0 = & \frac{1}{2 \gamma (\mode{P}{2}_0)^2}
  \label{eqn:fca_z0}
\end{align}
again parametrises the input pump power.
Note that negative values for $z_0$ are unphysical, because $\gamma$ is 
positive.
It is convenient to express lengths and powers in terms of the respective
natural units of $\Lambda$ (length) and $\Pi$ (power):
\begin{align}
  \Lambda = & \gamma / \PowerGain^2, & \Pi = & \PowerGain / \gamma.
  \label{eqn:dimless_units_fca}
\end{align}

Next, we insert \eqnref{eqn:fca_pump} into \eqnref{eqn:stokes_start} 
to obtain:
\begin{align}
  \frac{\dirderiv_z \mode{P}{1}}{\mode{P}{1}} = -\frac{1}{2 (z + z_0)}
  + \frac{\PowerGain}{\sqrt{2 \gamma (z + z_0)}},
\end{align}
The solution to this equation is
\begin{align}
  \mode{P}{1}(z) = & \frac{S^{\FCA}}{\sqrt{z + z_0}}
  \exp\left[ \PowerGain \sqrt{2 (z + z_0)/\gamma} \right],
  \label{eqn:ss_stokes_fca}
  \\
  \intertext{with the normalisation constant}
  S^{\FCA} = & \mode{P}{1}_0 \sqrt{z_0}
  \exp\left[ - \PowerGain \sqrt{2 z_0/\gamma} \right].
\end{align}
This solution \eqnref{eqn:ss_stokes_fca} is plotted in the top panel of
\figref{fig:sbs_fca_stokes} for $S^\FCA = \sqrt{\Lambda}$ directly above the 
inverse of \eqnref{eqn:ss_stokes_fca}, \ie the solution to the BSBS problem.
The first feature of \eqnref{eqn:ss_stokes_fca} is that the general shape of 
the solution is universal for SBS-waveguides with free carrier absorption.
Within this plot the waveguide corresponds a window starting at $z_0/\Lambda$
(defined by the pump power) and of length $L/\Lambda$.
Increasing the pump power simply moves the window to the left of the plot.
Variations in the SBS-gain, the FCA-coefficient and the injected Stokes power
only rescale the plot axes.
The second crucial feature of the solution is the extremum at $z + z_0 = \Lambda/2$. 
This means that in a strongly pumped waveguide the Stokes amplitude assumes
its minimum somewhere inside the waveguide and grows towards both ends.
Furthermore, for very high pump levels, the Stokes amplitude at the output is
lower than at the input, which means that any SBS-gain inside the waveguide is 
destroyed by free carrier absorption if the pump is too strong for a given
waveguide length.

\begin{figure}[t]
  \includegraphics[width=\columnwidth]{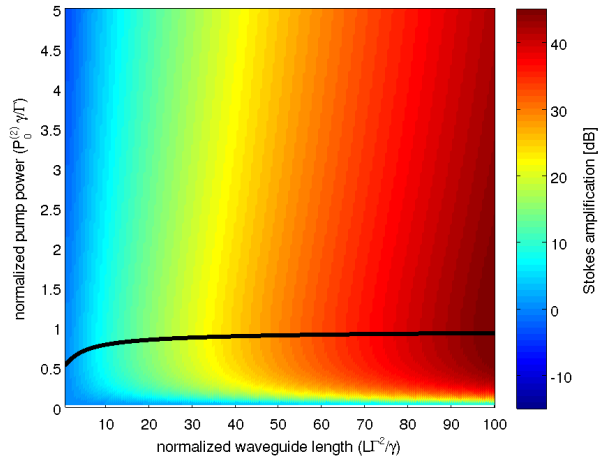}
  \caption{
    Stokes amplification in decibel of a waveguide whose only loss mechanism is
    2PA-induced free carrier absorption (2PA itself neglected) as a function
    of the pump power level and the waveguide length in natural units.
    Please note that for every fixed waveguide length, the amplification
    grows with increasing pump power, reaches a maximum and then drops off 
    again.
    These maxima correspond to optimal pump power levels and are connected with a
    solid black line.
  }
  \label{fig:ss_fca_gain}
\end{figure}

The Stokes amplification of the waveguide can be derived from 
\eqnref{eqn:ss_stokes_fca}; we find
\begin{align}
  \nonumber
  \TotalGain^{\FCA}(L) = & \frac{10}{\ln 10} \bigg\{
  \frac{\PowerGain}{\gamma \mode{P}{2}_0}
  \left[ \sqrt{2 \gamma L (\mode{P}{2}_0)^2 + 1} - 1 \right]
  \\
  & \quad
  - \frac{1}{2} \ln [ 1 + 2 \gamma L (\mode{P}{2}_0)^2]
  \bigg\}.
\end{align}
This function is shown in \figref{fig:ss_fca_gain} for a wide range of waveguide
lengths $L$ and pump powers $\mode{P}{2}_0$.
As in \figref{fig:sbs_fca_stokes}, this plot has been made universal for any 
combination of SBS and FCA coefficients.
As a consequence of the non-monotonic nature of \eqnref{eqn:ss_stokes_fca}, 
the Stokes amplification $\TotalGain^{\FCA}$ assumes a maximum at a specific 
pump power for every given waveguide length.
These optimal pump powers can be computed by solving the equation
\begin{align}
  \left[ 
    \frac{\partial \TotalGain^{\FCA}(L, \mode{P}{2}_0)}{\partial \mode{P}{2}_0} 
  \right]_L = 0
\end{align}
with $L$ kept fixed.
Since a closed analytical solution to this cannot be found, we use Newton's 
method to find the zeros.
However, it can be shown that the optimum pump powers always lies inside the 
interval $1/2 \leq \gamma^{-1}\PowerGain \mode{P}{2}_0 \leq 1$.
The numerically determined optimal pump powers are highlighted in 
\figref{fig:ss_fca_gain} with a solid black line.

\subsection{SBS with FCA and linear loss}
\label{sec:ss_fca_and_lin}

After the discussion of FCA as the only loss mechanism, we now add linear loss,
\ie we study the situation $\beta = 0$; $\{\alpha, \gamma\} \neq 0$.
This is the most general case involving FCA that still can be solved 
analytically.
In \secref{sec:ss_fca_lin_and_weak_tpa}, we present a perturbative treatment
of weak 2PA alongside strong linear loss and FCA based on the expressions 
derived in this section.

\begin{figure}[!t]
  \includegraphics[width=\columnwidth]{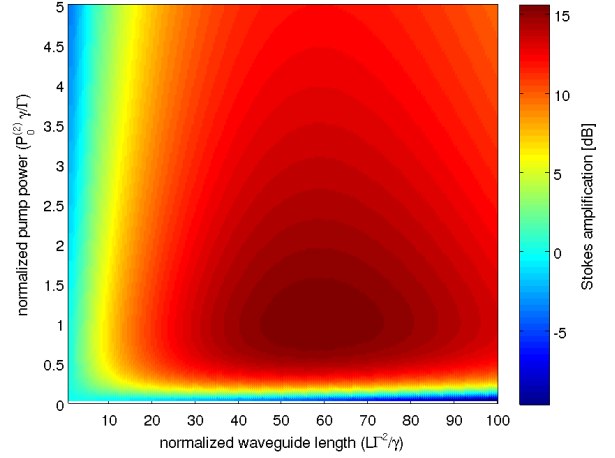}
  \caption{
    Stokes amplification in decibel of a waveguide that is subject to linear
    loss and 2PA-induced free carrier absorption (2PA itself neglected) as a 
    function of the pump power level and the waveguide length in natural units
    for a normalised linear loss 
    $\Upsilon = \alpha \gamma \PowerGain^{-2} = 0.03$.
    To illustrate the impact of linear loss, we adopted the colorbar from
    \figref{fig:ss_fca_gain}.
    There is one optimal combination of waveguide length 
    $L^{(\opt)} \approx 60 \gamma \Gamma^{-2}$ and pump power
    $P^{(\opt)} \approx \Gamma / \gamma$ that leads to the maximally
    realisable Stokes amplification of about $15 \text{ dB}$ for this value
    of $\Upsilon$, which corresponds to a figure of merit 
    (see \secref{sec:design_fca} for details) of $\FOM_\FCA \approx 2.9$.
  }
  \label{fig:ss_fca_lin_gain}
\end{figure}

As before, we start with solving the equation for the pump power along the 
waveguide:
\begin{align}
  \dirderiv_z \mode{P}{2} = -\alpha \mode{P}{2} -\gamma [\mode{P}{2}]^3.
\end{align}
This can be transformed into a Riccati equation via a substitution of the
type $u = [\mode{P}{2}]^2$.
The closed solution is:
\begin{align}
  \mode{P}{2}(z) = & \frac{\mode{P}{2}_0 \sqrt{\alpha}}
  {\sqrt{ \big( \gamma [\mode{P}{2}_0]^2 + \alpha \big) 
  \exp\big(2 \alpha z\big) - \gamma [\mode{P}{2}_0]^2} }. 
\end{align}
Next, we solve for the Stokes power along the waveguide by integrating
the equation
\begin{align}
  \frac{\dirderiv_z \mode{P}{1}}{\mode{P}{1}}
  = -\alpha + \PowerGain \mode{P}{2} - \gamma [\mode{P}{2}]^2,
\end{align}
to obtain
\begin{align}
  \nonumber
  \mode{P}{1}(z) = & S^{\FCA+\linear} \mode{P}{2}(z) \times
  \\ & \quad
  \exp\left\{ \frac{\PowerGain}{\sqrt{\alpha \gamma}} \tan^{-1}\left[
  \frac{\sqrt{\alpha}}{\sqrt{\gamma}\mode{P}{2}(z)} \right] \right\},
  \label{eqn:ss_stokes_fca_and_lin}
\end{align}
where the input Stokes power enters via the normalisation constant
\begin{align}
  S^{\FCA+\linear} = & \frac{\mode{P}{1}_0}{\mode{P}{2}_0} 
  \exp\left\{ -\frac{\PowerGain}{\sqrt{\alpha \gamma}} \tan^{-1}\left[
  \frac{\sqrt{\alpha}}{\sqrt{\gamma}\mode{P}{2}_0} \right] \right\}.
  \label{eqn:ss_stokes_fca_and_lin_norm}
\end{align}
The spatial evolution of the Stokes wave is qualitatively similar to the one
depicted in \figref{fig:sbs_fca_stokes}.
By taking the decadic logarithm of \eqnref{eqn:ss_stokes_fca_and_lin} including 
\eqnref{eqn:ss_stokes_fca_and_lin_norm}, one can directly obtain the explicit 
relationship between the total Stokes amplification $\TotalGain^{\FCA+\linear}$ 
and the waveguide length and the pump power.
This expression is long and convoluted and does not provide any additional 
insight, so we forgo showing it here.
As before, it is very useful to introduce the problem-specific units of length 
and power stated in \eqnref{eqn:dimless_units_fca} as the natural unit system.
Consequently, the linear loss is best expressed in natural units of inverse 
length: 
\begin{align}
  \Upsilon = \alpha \gamma \PowerGain^{-2}.
\end{align}
If this normalised linear loss vanishes, the plot of $\TotalGain^{\FCA+\linear}$ 
is identical to the previous result \figref{fig:ss_fca_gain}.
As the linear loss increases, the total amplification decays not only for 
increasing pump power but also for increasing waveguide length, leading to a 
well defined maximal obtainable amplification for each value $\Upsilon$.
The maximal amplification is obtained for exactly one pair of optimal pump
power $P^{(\opt)}$ and optimal waveguide length $L^{(\opt)}$.
This is illustrated in \figref{fig:ss_fca_lin_gain} for the case of 
$\Upsilon = 0.03$.
We will resume this topic in \secref{sec:design_fca}.

\subsection{SBS with FCA, linear loss and weak 2PA}
\label{sec:ss_fca_lin_and_weak_tpa}

Finally, we turn towards the case that all three loss mechanisms (linear loss,
2PA and FCA) are present.
In this case, the equation for $\mode{P}{2}(z)$ can only be derived implicitly,
\ie in the form $z(\mode{P}{2})$, but not explicitly.
Unfortunately, this makes it impossible to solve the general case exactly.
One therefore has to resort to approximate solutions derived from numerical
calculations or perturbation theory in the parameter $\beta$.
To zeroth order in $\beta$, the pump profile $\mode{P}{2}(z)$ can be adopted
from the 2PA-free case \eqnref{eqn:fca_pump}. 
This is because we assume that the impact of 2PA on the pump is of the same order
as the (neglected) pump-depletion due to SBS.
Under this assumption, the SBS-gain $\PowerGain$ is simply reduced by twice the 
2PA-coefficient:
\begin{align}
  \PowerGain_\eff = \PowerGain - 2\beta.
\end{align}
All solutions from \secref{sec:ss_fca_and_lin} apply directly with this
substitution.

The next perturbation order would involve a first order correction to the pump
profile $\mode{P}{2}(z)$. 
While this can be found, the subsequent equation for the Stokes amplitude
$\mode{P}{1}(z)$ and as a consequence the equation for the total Stokes 
amplification involve an integral that does not have a closed solution.
For this reason, we adhere to the zeroth order approximation and refer the
reader to numerical solutions whenever more accurate results are required.

\section{Design guidelines and figures of merit}
\label{sec:design}

In this section, we provide guidelines for the optimal choice of waveguide 
length and pump power of waveguides whose SBS gain and loss coefficients have
either been measured or computed numerically using the expressions in 
Ref.~\cite{Wolff2015a} and the Appendices of this paper.
We furthermore provide figures of merit that express the 
suitability of any given material or waveguide design for the purpose of SBS
in a single number.

\subsection{SBS with 2PA and linear loss}
\label{sec:design_tpa}

In the absence of fifth-order loss (especially 2PA-induced FCA),
we can apply the results from \secref{sec:ss_tpa_and_lin}.
This covers most centrosymmetric and amorphous insulators including glasses.
The main feature of \eqnref{eqn:ss_gain_tpa_and_lin} is that it is monotonic
and unbounded with respect to the pump power.
Thus, the SBS gain of any given waveguide can in principle be increased 
indefinitely by injecting a sufficiently strong pump, provided the SBS gain is
sufficient to overcome the loss due to 2PA.
In other words, any material or waveguide design is capable of exhibiting
SBS if the figure of merit
\begin{align}
  \FOM_\TPA = \frac{\PowerGain}{2 \beta}
  \label{eqn:design_fom_tpa}
\end{align}
is greater than one.
In every material or waveguide design with $\FOM_\TPA < 1$, SBS is 
quenched by 2PA.
For $\FOM_\TPA = 1$, SBS and 2PA cancel each other to leading order;
higher order corrections predict a weak power-dependent decay of the Stokes 
wave in this regime.

The total Stokes amplification does not grow indefinitely with respect to 
waveguide length, because the linear term $-\alpha L$ in 
\eqnref{eqn:ss_gain_tpa_and_lin} at some point overcomes the saturating 
logarithm of the second term.
The optimal waveguide length $L^{(\opt)}$ is given by the condition
\begin{align}
  \frac{\partial \TotalGain^{\TPA+\linear}}{\partial L} = 0,
\end{align}
which can be evaluated exactly:
\begin{align}
  L^{(\opt)} = & \frac{1}{\alpha} \ln \left[
    \frac{\mode{P}{2}_0 (\PowerGain - \beta)}{(\beta \mode{P}{2}_0 + \alpha)}
  \right]
  \\
  & \ \xrightarrow{\mode{P}{2}_0 \gg \alpha / \beta}  \ 
  \frac{\ln ( 2 \FOM_\TPA - 1 )}{\alpha} .
\end{align}
This means that the optimal waveguide length depends logarithmically on the 
pump power, but will always remain in the vicinity of $\alpha^{-1}$.
The power-dependence of the optimal waveguide length for the case of 
$\FOM_\TPA = 10$ is shown in \figref{fig:ss_tpa_gain} as a black solid line.

\subsection{SBS with FCA, linear loss and weak 2PA}
\label{sec:design_fca}

In most indirect semiconductors such as silicon and germanium, the long carrier 
lifetime leads to the effect that the free carriers created by 2PA have a much
stronger impact on the absorption of a quasi-CW light wave than the two-photon
absorption itself.
Although this effect can be reduced by extracting free carriers via an 
externally applied electric field, the loss due to 2PA-induced FCA will still 
surpass the 2PA itself in many situations and the results from 
\secref{sec:ss_fca_lin_and_weak_tpa} can be applied.

In contrast to the previous section, the total Stokes amplification of a 
waveguide that experiences FCA is non-monotonic with power and is in fact 
bounded with respect to 
both the injected pump power and its length.
For each set of SBS gain and loss parameters, there is exactly one choice of 
length and pump power that leads to the maximal gain, as can be seen
in \figref{fig:ss_fca_lin_gain}.

In any given material or waveguide design, a total Stokes amplification can be
obtained only if the FCA figure of merit
\begin{align}
  \FOM_\FCA = \frac{\PowerGain - 2 \beta}{2 \sqrt{\alpha \gamma}}
  \label{eqn:design_fom_fca}
\end{align}
is greater than 1.
As shown in~\cite{Wolff2015b}, this quantity emerges naturally as a figure
of merit from Eqs.~\eqref{eq:governing}.
Unlike the situation without FCA (see \secref{sec:design_tpa}), $\FOM_\FCA$
merit also implies an absolute maximum for the Stokes amplification that can be 
obtained with a given waveguide design.
It is attained for the optimal choice of waveguide length and pump power, both 
of which depend on $\FOM_\FCA$.
This interdependence is discussed in a separate paper~\cite{Wolff2015b}.

\begin{figure} 
  \includegraphics[width=\columnwidth]{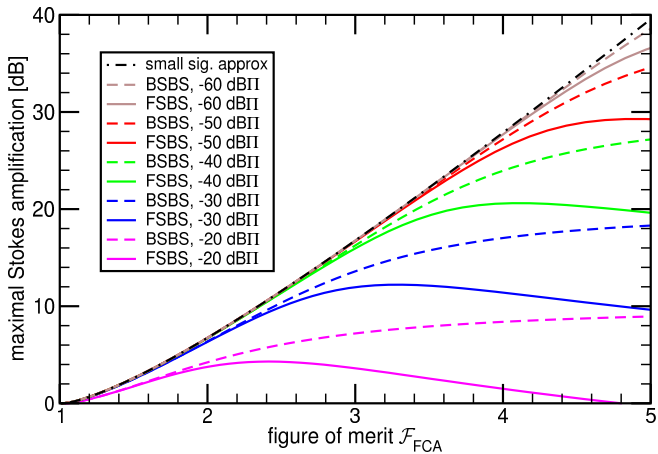}
  \caption{
    Stokes amplification plots for linear loss, FCA and weak 2PA:
    Comparison of the maximally realisable amplification within the 
    small-signal approximation (black dash-dotted line) and corresponding 
    maximal amplifications in FSBS (solid lines) and BSBS (dashed lines)
    configuration for finite Stokes input powers between 
    $-60\,\text{dB}$ and $-20\,\text{dB}$ of the natural power unit 
    $\Pi = \PowerGain_\eff / \gamma = (\PowerGain - 2 \beta) / \gamma$.
  }
  \label{fig:design_fca_comp}
\end{figure}
In \figref{fig:design_fca_comp}, we compare the maximally realisable Stokes 
amplification predicted by our small-signal approximation with results obtained 
by numerically solving the large-signal equations 
Eqs.~(\ref{eqn:sbs_stokes_pwr}, \ref{eqn:sbs_pump_pwr}) for the analytically 
predicted operating conditions and injected Stokes powers between 
$-60\,\text{dB}$ and $-20\,\text{dB}$ of the natural power unit 
$\Pi$ both in forward and backward configuration.
In the case of FSBS, we employed a simple 4-th order Runge-Kutta integrator.
In the case of BSBS, we used the shooting method based on a variable order
Runge-Kutta that is optimised for stiff differential equations.
It is clearly visible how the small-signal solution provides an upper
bound for the realisable Stokes amplification.
The difference can be attributed to energy transfer from the pump to the Stokes
(known as pump depletion in SBS without loss) and additional carrier generation
due to the relatively high Stokes intensity.
Finally, we note that the optimal operating conditions (\ie waveguide length 
and pump power) shift as the input Stokes level is increased and that this
affects forward SBS and backward SBS quite differently.
This is because the small-signal approximation no longer applies over the full
range of $\FOM_\FCA$ and input Stokes powers.
In the case of BSBS, the amplification maximum as a function of waveguide length
and pump power remains very flat, which means that the analytical prediction is
very viable.
However, in the case of FSBS the amplification maximum becomes fairly sharp
and the waveguide has to be shortened to obtain performance that is comparable
to the BSBS amplification shown in \figref{fig:design_fca_comp}.
In other words: A forward SBS amplifier has a smaller dynamic range than an
amplifier based on backward SBS.

\section{Conclusions}
\label{sec:conclusion}

In this paper, we have discussed the impact of nonlinear optical loss on the
process of SBS within a small-signal approximation.
Based on analytical solutions to this problem, we have derived figures of merit
that describe the suitability of a material or a waveguide design for the case
that fifth-order nonlinear loss effects such as 2PA-induced FCA can be neglected
and the case that they cannot.
In the former case, we find that, although third-order loss reduces the total
Stokes wave amplification along the waveguide, the amplification is not 
fundamentally restricted and can be increased indefinitely by increasing pump
power until the weakest component in the optical circuit is permanently damaged.
In the latter case, however, the Stokes amplification due to SBS is 
overcome by FCA once some optimal pump power is exceeded.
Thus, the total amplification is bounded to a value that is ultimately given by
the figure of merit $\FOM_\FCA = (\PowerGain - 2 \beta) / \sqrt{4 \alpha \gamma}$.
This opens basically three possible routes to effective SBS-circuits in 
semiconductor (especially silicon) photonics platforms.
First, the upper amplification bound can be increased by reducing the linear
loss $\alpha$.
This often is a challenging task.
Second, the impact of FCA could be drastically reduced by removing the free
carriers by means of an externally applied electric field.
The resulting increase in linear loss would be tolerable as long as the 
product $\alpha \gamma$ is reduced.
The third, and on a short time scale potentially most viable, solution would be 
to design future silicon photonic circuits for SBS to operate at wavelengths 
below the 2PA-threshold, \eg around $2400\,\text{nm}$ for the case of silicon.
This would eliminate not only 2PA (which itself is not problematic), but in 
particular the induced FCA and lead to higher acoustic quality factors due to
the reduced acoustic frequency.

\section*{Acknowledgements}
We acknowledge financial support from the Australian Research Council (ARC)
via the Discovery Grant DP130100832, its Laureate Fellowship
(Prof. Eggleton, FL120100029) program and the ARC Center of Excellence CUDOS
(CE110001018).

\begin{appendix}
\begin{widetext}

\section{Expressions for loss coefficients}
\label{appx:loss_coeffs}

In the following Appendices, we derive the loss coefficients that appear in 
\eqnref{eqn:sbs_stokes} and \eqnref{eqn:sbs_pump}.
The explicit expressions are:
\begin{align}
  \tilde \alpha_i = &\frac{\eps_0 \omega}{\mode{\Power}{i}} \int \total^2 r \ 
  \big| \mode{\widetilde{\vec e}}{i} \big|^2 \Im\{\eps_r\},
  \label{eqn:appx_linear}
  \\
  \tilde \beta_{ij} = & 
  \frac{1}{\mode{\Power}{i}} \int \total^2 r \ \big(
  \big| \mode{\widetilde{\vec e}}{i} \cdot \mode{\widetilde{\vec e}}{j} \big|^2 + 
  \big| \mode{\widetilde{\vec e}}{i} \cdot (\mode{\widetilde{\vec e}}{j})^\ast \big|^2 + 
  \big| \mode{\widetilde{\vec e}}{i} \big|^2 \big| \mode{\widetilde{\vec e}}{j} \big|^2 \big)
  \Sigma^\TPA,
  \label{eqn:appx_tpa}
  \\
  \tilde \gamma_{ijk} = & \frac{1}{\mode{\Power}{i}}
  \int \total^2 r \ |\mode{\widetilde{\vec e}}{i}|^2
  \Big[ |\mode{\widetilde{\vec e}}{j} \cdot \mode{\widetilde{\vec e}}{k}|^2 + 
    |\mode{\widetilde{\vec e}}{j} \cdot (\mode{\widetilde{\vec e}}{k})^\ast|^2 +
    |\mode{\widetilde{\vec e}}{j}|^2 |\mode{\widetilde{\vec e}}{k}|^2 
  \Big]
  \Sigma^\FCA,
  \label{eqn:appx_fca}
\end{align}
where the indices $i$, $j$ and $k$ label the respective optical eigenmodes 
modes and can take the values 1 and 2.
The Symbols $\Sigma^\TPA$ and $\Sigma^\FCA$ are nonlinear conductivities 
associated with 2PA and 2PA-induced FCA and are expressed in more conventional
terms in \eqnref{eqn:appx_sigma_tpa} and \eqnref{eqn:appx_sigma_fca}, 
respectively.
The values
\begin{align}
  \Sigma^\TPA = & 5.5 \cdot 10^{-16} \, \text{W} \text{m}^3\text{V}^{-4},
  &
  \Sigma^\FCA = & 6.0 \cdot 10^{-28} \, \text{W} \text{m}^5\text{V}^{-6},
\end{align}
correspond to a bulk 2PA-coefficient of $5 \cdot 10^{-12} \text{ m/W}$ and
an electron scattering cross section of $1.45 \cdot 10^{-21} \text{ m}^2$
together with a carrier life time of $10\,\text{ns}$.
Those are typical literature values~\cite{Jalali2005} for silicon at a vacuum 
wavelength of $1550\text{ nm}$.

\section{Derivation of 2PA-terms}
\label{appx:tpa}

In time-domain it is sometimes advantageous to represent lossy (\ie dispersive) 
optical nonlinearities is via a nonlinear current 
distribution~\cite{Taflove2005}.
The corresponding nonlinear conductivity to represent 2PA is related to the
imaginary part of the third-order susceptibility:
\begin{align}
  \Sigma^{\TPA} = \omega \eps_0 \Im\{\chi^{(3)}\}.
  \label{eqn:appx_sigma_tpa}
\end{align}
Here, we neglected the tensorial nature of $\chi^{(3)}$ in order to to improve 
the readability of the integrals (\ref{eqn:appx_linear}--\ref{eqn:appx_fca}). 
The generalisation to tensorial nonlinearities is a matter of book-keeping.
With this, we find the real-valued time-domain current density
$
  \vec J^\TPA(t) = - \Sigma^\TPA |\vec E(t)|^2 \vec E(t),
$
where for the sake of brevity, we wrote $\vec v^2 = \vec v \cdot \vec v$ and
$|\vec v|^2 = \vec v^\ast \cdot \vec v$ for any vectorial quantity $\vec v$.
The electric field is formed by the interference between two optical
eigenmodes:
$
  \vec E(t) = \mode{a}{1} \mode{\vec e}{1} + \mode{a}{2} \mode{\vec e}{2} + \cc.
  $
Within a coupled-mode theory, we need the projection of $\vec J^\TPA$ on the 
optical eigenmodes, \eg $\mode{\vec e}{1}$:
\begin{align}
  \langle \mode{e}{1} | \vec J^\TPA \rangle 
  = & -\acaverage{\int \total^3 r \ \left[ (\mode{\vec e}{1})^\ast \cdot 
  (\mode{a}{1} \mode{\vec e}{1} + \mode{a}{2} \mode{\vec e}{2} + \cc) \right]
  \Sigma^\TPA \big|\mode{a}{1} \mode{\vec e}{1} + 
  \mode{a}{2} \mode{\vec e}{2} + \cc \big|^2}
  \\
  \nonumber
  = & -\Big\langle
  2 \mode{a}{1} \int \total^3 r \ \big|\mode{\vec e}{1}\big|^2
  \Sigma^\TPA \left[ \big| \mode{a}{1}\big|^2 \big| \mode{\vec e}{1}\big|^2 
    + \big| \mode{a}{2}\big|^2 \big| \mode{\vec e}{2}\big|^2 \right]
    \\
    \nonumber
    & \quad + \big(\mode{a}{1}\big)^\ast \int \total^3 r \ 
    \big(\mode{\vec e}{1} \cdot \mode{\vec e}{1} \big)^\ast \Sigma^\TPA 
    \big( \mode{a}{1} \big)^2
    \mode{\vec e}{1} \cdot \mode{\vec e}{1}
    \\
    \nonumber
    & \quad + 2 \mode{a}{2} \int \total^3 r \ 
    \big(\mode{\vec e}{1}\big)^\ast \cdot \mode{\vec e}{2} \Sigma^\TPA 
    \mode{a}{1} \big( \mode{a}{2} \big)^\ast
    \mode{\vec e}{1} \cdot \big(\mode{\vec e}{2}\big)^\ast
    \\
    \nonumber
    & \quad 
    + 2 \big(\mode{a}{2}\big)^\ast \int \total^3 r \ 
    \big(\mode{\vec e}{1} \cdot \mode{\vec e}{2}\big)^\ast \Sigma^\TPA 
    \mode{a}{1} \mode{a}{2} \mode{\vec e}{1} \cdot \mode{\vec e}{2}
  \\
  & \quad + \text{oscillating terms} \Big \rangle_{T_\acoustic}
  \\
  \nonumber
  = & \mode{a}{1} \big| \mode{a}{1} \big|^2 \ \int \total^2 r \ \Sigma^\TPA \big(
    2 \big| \mode{\widetilde{\vec e}}{1} \cdot (\mode{\widetilde{\vec e}}{1})^\ast \big|^2 + 
    \big| \mode{\widetilde{\vec e}}{1} \cdot \mode{\widetilde{\vec e}}{1} \big|^2 \big)
  \\
  & \quad + 2 \mode{a}{1} \big| \mode{a}{2} \big|^2 \ \int \total^2 r \ \Sigma^\TPA \big(
    \big| \mode{\widetilde{\vec e}}{1} \cdot \mode{\widetilde{\vec e}}{2} \big|^2 + 
    \big| \mode{\widetilde{\vec e}}{1} \cdot (\mode{\widetilde{\vec e}}{2})^\ast \big|^2 + 
    \big| \mode{\widetilde{\vec e}}{1} \big|^2 \big| \mode{\widetilde{\vec e}}{2} \big|^2 \big).
\end{align}
The projection on the other mode $\mode{\vec e}{2}$ follows from this by 
interchanging the mode index superscripts.

\section{Derivation of FCA-terms}
\label{appx:fca}

As in Appendix~\ref{appx:tpa} we intend to describe the loss via a time domain current
density $\vec J^\FCA$.
It is due to the conductivity of a dilute plasma with carrier density 
$N_\text{c}$, which can be described~\cite{Tinten2000} using a Drude-Sommerfeld 
model with effective carrier mass $m^\ast$ and damping parameter 
$\omega_\text{d}$:
\begin{align}
  \vec J^\FCA(t) = - \frac{\omega_\text{d} q^2 N_\text{c}}
  {(\omega^2 + \omega^2_\text{d}) m^\ast} \vec E(t).
\end{align}
Here, $q$ is the elementary charge and the parameters $m^\ast$, 
$\omega_\text{d}$ are material specific and therefore may depend on position 
within the waveguide.
The carriers are created by two-photon absorption and destroyed via recombination
on a time scale $\tau_\text{c} > 10\text{ns}$.
The order of this time constant is important because it covers around $100$ 
acoustic cycles for typical Stokes shifts of $3-30\text{GHz}$.
Consequently, the carrier density is to be derived from the 2PA power loss 
averaged over a time-scale greater than one acoustic period:
\begin{align}
  P^\TPA_\loss = & -\acaverage{\vec E(t) \cdot \vec J^\TPA(t)}
  = \Sigma^\TPA \acaverage{|\vec E(t)|^2 |\vec E(t)|^2}
  \\
  \nonumber
  = & 2 \Sigma^\TPA \bigg\{
  |\mode{a}{1}|^4 
  \Big[|\mode{\vec e}{1} \cdot \mode{\vec e}{1}|^2 + 2 (|\mode{\vec e}{1}|^2)^2 \Big]
  \ + \ |\mode{a}{2}|^4 
  \Big[|\mode{\vec e}{2} \cdot \mode{\vec e}{2}|^2 + 2 (|\mode{\vec e}{2}|^2)^2 \Big]
  \\
  & \quad
  + 4 |\mode{a}{1}|^2 |\mode{a}{2}|^2
  \Big[ |\mode{\vec e}{1} \cdot \mode{\vec e}{2}|^2 + 
    |\mode{\vec e}{1}|^2 |\mode{\vec e}{2}|^2 + 
    |\mode{\vec e}{1} \cdot (\mode{\vec e}{2})^\ast|^2
  \Big]
  \bigg\}.
\end{align}
The carrier generation rate is this optical power loss divided by twice the
photon energy, whereas the carrier recombination rate is determined by $N_c$
itself and the carrier life time $\tau_c$:
$  \partial_t N_c = P^\TPA_\loss / (2 \hbar \omega) - \frac{N_c}{\tau_c}. $
Note that the carrier life typically is reduced near material interfaces
and therefore position dependent.
In our crude model we neglected carrier diffusion.
If the field amplitudes vary on time scales larger than $\tau_c$, the loss
due to FCA is given by the equilibrium carrier density 
$
N_c^{\text{(eq)}} = \tau_c P^\TPA_\loss / (2 \hbar \omega).
$
This allows us to express the FCA-related nonlinear conductivity:
\begin{align}
  \Sigma^\FCA = \frac{\omega_\text{d} \tau_\text{c} q^2 \Sigma^\TPA}
  {2 \hbar \omega (\omega^2 + \omega^2_\text{d}) m^\ast}.
  \label{eqn:appx_sigma_fca}
\end{align}
Within the context of our coupled mode theory, we require the projection of the
FCA-current on the optical eigenmodes, \eg the mode $\mode{\vec e}{1}$:
\begin{align}
  \langle \mode{e}{1} | \vec J^\FCA \rangle 
  = & - \frac{1}{2\hbar \omega} 
  \optaverage{ \int \total^3 r \ [ \mode{\vec e}{1} ]^\ast
    \cdot [\mode{a}{1} \mode{\vec e}{1} + \mode{a}{2} \mode{\vec e}{2} + \cc ]
    \frac{\omega_\text{d} q^2 \tau_\text{c}}
    {2 \hbar \omega (\omega^2 + \omega_\text{d}^2) m^\ast} 
  P^\TPA_\loss }
  \\
  \nonumber
  = & -
  \frac{\mode{a}{1} |\mode{a}{1}|^4 }{\hbar \omega}
    \int \total^2 r \ |\mode{\widetilde{\vec e}}{1}|^2
  \Sigma^\FCA
  \Big[|\mode{\widetilde{\vec e}}{1} \cdot \mode{\widetilde{\vec e}}{1}|^2 
  + 2 (|\mode{\widetilde{\vec e}}{1}|^2)^2 \Big]
  \\
  \nonumber
  & \quad -
  \frac{\mode{a}{1} |\mode{a}{2}|^4 }{\hbar \omega}
    \int \total^2 r \ |\mode{\widetilde{\vec e}}{1}|^2
  \Sigma^\FCA
  \Big[|\mode{\widetilde{\vec e}}{2} \cdot \mode{\widetilde{\vec e}}{2}|^2 
  + 2 (|\mode{\widetilde{\vec e}}{2}|^2)^2 \Big]
  \\
  & \quad - 
  \frac{4 \mode{a}{1} |\mode{a}{1}|^2 |\mode{a}{2}|^2 }{\hbar \omega}
    \int \total^2 r \ |\mode{\widetilde{\vec e}}{1}|^2
  \Sigma^\FCA
  \Big[ |\mode{\widetilde{\vec e}}{1} \cdot \mode{\widetilde{\vec e}}{2}|^2 + 
    |\mode{\widetilde{\vec e}}{1}|^2 |\mode{\widetilde{\vec e}}{2}|^2 + 
    |\mode{\widetilde{\vec e}}{1} \cdot (\mode{\widetilde{\vec e}}{2})^\ast|^2
  \Big].
\end{align}

\end{widetext}

\end{appendix}


\end{document}